\documentclass[11pt]{article}
\usepackage{moriond2000,epsfig}

\def\be{\begin{equation}}
\def\ee{\end{equation}}
\def\bea{\begin{eqnarray}}
\def\eea{\end{eqnarray}}

\newcommand{\ltapprox}{\raisebox{-0.5ex}{$\,\stackrel{<}{\scriptstyle\sim}\,$}}
\newcommand{\gtapprox}{\raisebox{-0.5ex}{$\,\stackrel{>}{\scriptstyle\sim}\,$}} 

\begin{document}
\vspace*{4cm}
\title{PHOTOMETRIC REDSHIFTS AND GRAVITATIONAL TELESCOPES}

\author{R. PELLO$^1$, M. BOLZONELLA$^{2,1}$, J.F. LE BORGNE$^1$, J.P. KNEIB$^1$,
B. FORT$^3$, Y. MELLIER$^3$, \\
M. DANTEL$^4$, L. CAMPUSANO$^5$, R.S. ELLIS$^6$, I. SMAIL$^7$,
I. TIJERA$^8$}

\address{
$^1$ LAT, Observatoire Midi-Pyr\'en\'ees, UMR5572, 14 Av. Edouard-B\'elin,
31400 Toulouse, France \\
$^2$ Istituto di Fisica Cosmica ``G. Occhialini'', via Bassini 15, I-20133 Milano, Italy \\
$^3$ Institut d'Astrophysique de Paris, 98 bis boulevard Arago, 75014 Paris, France \\
$^4$ Observatoire de Paris, DEMIRM, 61 Av. de l'Observatoire, 75014 Paris, France \\
$^5$ Observatorio Cerro Cal\'an, Dept. de Astronom\' \i a, U. de Chile, Casilla 36-D, Santiago, Chile \\
$^6$ Astronomy 105-24, Caltech, Pasadena CA 91125, USA \\
$^7$ Department of Physics, University of Durham, South Road, Durham DH1 3LE, England \\
$^8$ Departament d'Astronomia i Meteorologia, U. de Barcelona, Diagonal 648, 08028 Barcelona, Spain
}

\maketitle\abstracts{We review the use of photometric redshifts in the particular 
context of Gravitational Telescopes. We discuss on the possible application
of such a technique to the study of both the faint population of distant sources and the
properties of
cluster lenses. Photometric reshifts could be used to derive the redshift distribution and 
properties of a very faint subsample of high-$z$ lensed galaxies, otherwise out of reach.
Concerning the mass distribution in lensing clusters, photometric reshifts are
strongly needed to scale the mass in weak lensing analysis, and to characterize the
lensing structure.
}

\section{Introduction}

Photometric redshifts (hereafter $z_{phot}$s) are a promising technique in deep 
universe studies. The interest for this technique has recently increased with the 
development of large field and deep field surveys, in particular the HDF. The 
relatively high number of objects accessible to photometry allows to enlarge the 
spectroscopic sample towards the faintest magnitudes. For this reason, $z_{phot}$s
are also important when using lensing clusters as Gravitational Telescopes (GTs),
for the study of both the background sources and the lenses. 

   One of the most widely used $z_{phot}$ techniques is the SED fitting procedure.
The observed photometric spectral energy distributions (SEDs) are compared to those obtained
from a set of template spectra, using the same photometric system.
The method is based on the detection of strong spectral features, and it 
has been largely applied on HDF studies \cite{Mob} \cite{Lan} \cite{Gwy}
\cite{Saw} \cite{Gia} \cite{Fdez} \cite{Arn} \cite{Fur}.
The examples presented in this paper have been produced using our public code 
called {\it hyperz\/}, which adopts a standard SED fitting method, but most 
results should be completely general when using other $z_{phot}$ tools.
{\it hyperz\/} is presented in a recent paper by Bolzonella et al. (2000) \cite{Bol},
and it is available on the web at {\tt http://webast.ast.obs-mip.fr/hyperz\/}.
When applying {\it hyperz} to the spectroscopic samples available on the HDF,
the uncertainties are typically $\delta z / (1 + z) \sim 0.1$, and this value
gives an idea of the expected accuracy of $z_{phot}$s for the purposes of this
paper.

\section{Photometric redshifts and background sources}

The idea is to take benefit from the amplification factor in GTs 
to study the properties of the distant background population of lensed galaxies.
The typical amplification ranges between 1 to 3 magnitudes in the cluster core.
In principle, GTs are useful to build-up and study an independent sample of high-$z$ galaxies,
which complements the large samples obtained in standard field surveys. The 
advantage is that this sample is less biased in luminosity than the field.

\subsection{Redshift distribution of faint galaxies}

One of the main goals of the GT is to determine the $z$ distribution
of a very faint subsample of high-$z$ lensed galaxies, invisible otherwise, and
this can be achieved using $z_{phot}$s.
In order to prevent the biases towards or against a particular type of galaxy or 
redshift domain, $z_{phot}$ shall be computed from broad-band photometry
using a large wavelength interval, from B (U when possible) to K. This allows also
to reduce the errors on $z_{phot}$ (see Bolzonella et al. 2000 for a detailed discussion).

   We have obtained the (photometric) N($z$) distribution of arclets in several well 
known clusters (A2390, A370, Cl2244-02, AC114,...). 
Figure~1 displays a recent example, the
$z_{phot}$  distribution for different source samples in MS1008-1224\cite{Athreya},
corresponding to different limits in magnitude. In this case, $z_{phot}$s were computed
from  VLT BVRI (FORS) and JK' (ISAAC) public data obtained during the Science Verification phase.
The sample includes 559 sources located on the central 2.5 arcminute field of ISAAC,
excluding obvious cluster members.
The typical number of high-$z$ sources found in the inner 1' radius region of the
cluster is $\sim 30$ to 50 at $1 \le z \le 7 $, for a photometric survey performed 
with the HST and different 4m class telescopes. 
Clusters with well constrained mass distributions enable to recover precisely the 
$N(z_{phot})$ distribution of lensed galaxies, by correcting the relative impact
parameter on each redshift bin. 

\begin{figure}
\hbox{
\psfig{figure=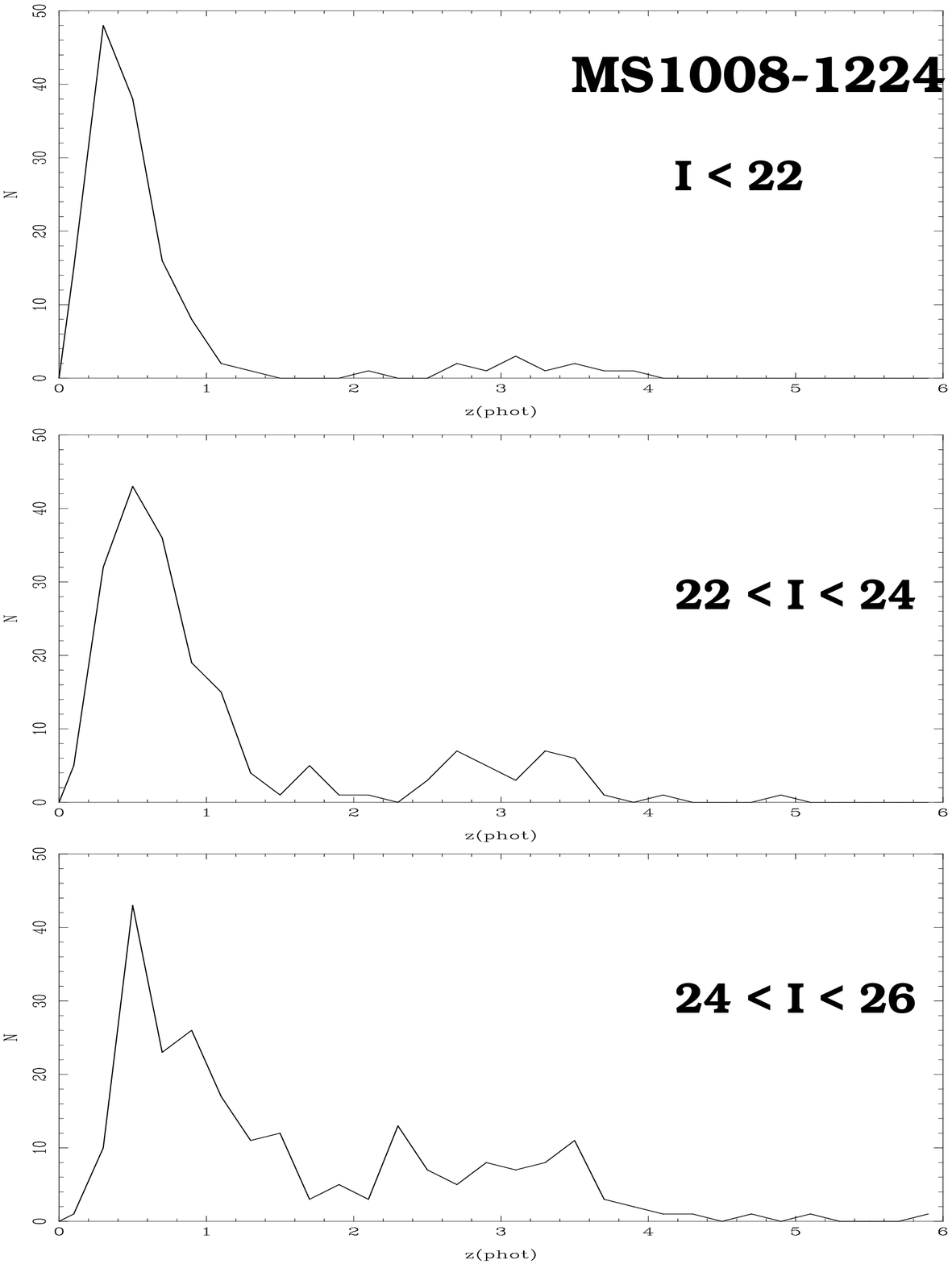,height=10.0cm}
\psfig{figure=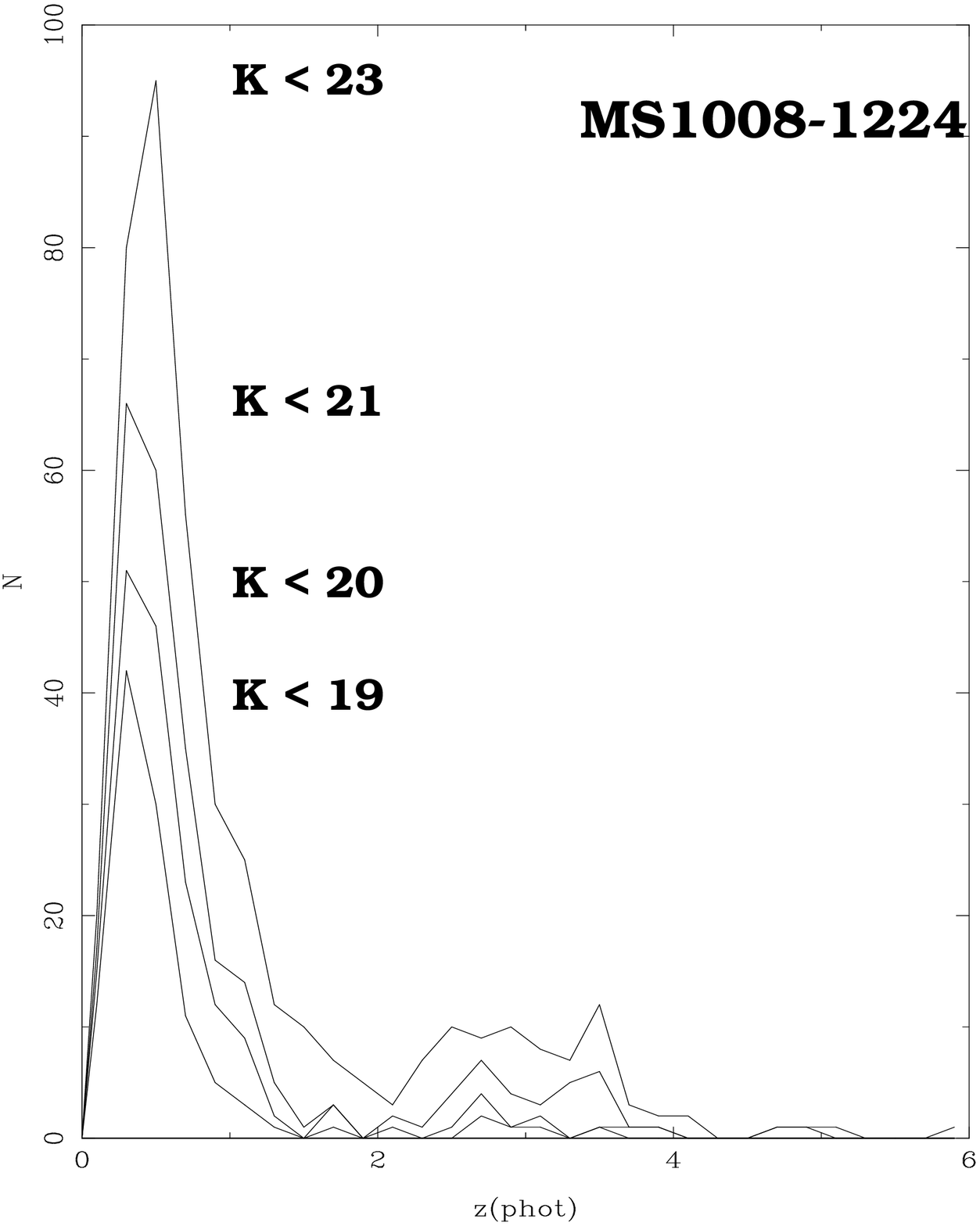,height=10.0cm}
}
\caption{Photometric redshift distribution of 559 gravitationally amplified 
sources in the core of MS1008-1224. $z_{phot}$s were obtained from BVRIJK'
photometry. Different limits in magnitude are considered in I (left) and K'
(right) bands.
\label{fig1}}
\end{figure}

\subsection{Optimization of spectroscopic redshift surveys}

An interesting issue for $z_{phot}$ when using GTs for the spectroscopic study 
of faint amplified sources is the optimization of the survey, that is 
selecting the best spectral domain in the visible or near-IR bands. 
This means in practice to produce a criterion based in $z_{phot}$ to discriminate 
between objects showing strong spectral features in the optical and in the 
near-IR. 

An additional benefit of $z_{phot}$ is that this technique efficiently contributes
to the identification of objects with ambiguous spectral features,
such as single emission lines. An example of this is given in a recent paper 
by Campusano et al. (in preparation, see also Le Borgne et al., this conference).

\subsection{Identification and study of very high-$z$ sources}

The signal/noise ratio in spectra of amplified sources
and the detection fluxes are improved beyond the limits of conventional techniques,
whatever the wavelength used for this exercise. An example is the 
ultra-deep MIR survey of A2390\cite{Alt}, and the SCUBA
cluster lens survey\cite{Smail}\cite{Blain}. Number of $z \gtapprox 4$ lensed 
galaxies have been found recently, and these findings strongly encourage 
our approach\cite{Tra}, \cite{Franx}, \cite{Fry}, \cite{pel2}.
Cluster lenses are the natural way to search for primeval galaxies, in order to
constraint the scenarios of galaxy formation.

High-$z$ lensed sources could be selected close to the appropriate critical lines,
and identified using $z_{phot}$ criteria. $z_{phot}$s are computed from broad-band 
photometry on a large wavelength interval, from B (U when possible) to K. 

Whatever the $z_{phot}$ method used, a crucial test is the comparison 
between the photometric and the spectroscopic redshifts obtained on a
restricted subsample of objects. Thanks to the magnification factor, 
cluster lenses could be used to enlarge the training sample towards 
the faintest magnitudes.

\subsection{Combining photometric and lensing redshifts}

Lensing inversion and $z_{phot}$ techniques produce independent probability
distributions for the redshift of amplified sources. Therefore, the combination of 
both methods provides an alternative way 
to determine the redshift distribution of the faintest high-$z$ sources.
Figure~2 displays the results for 4 sources in A2390,
with $z_{phot}$ computed from BgVrRiIJK' photometry\cite{pel2}.
When comparing the $z_{phot}$ and lensing redshift ($z_{lens}$) values for a subsample of 98 arclets
in the core of A2390, all selected according to morphological criteria (minimum elongation and 
right orientation are requested), we find that about $60\%$ of the sample 
have $|z_{phot} - z_{lens}| \le 0.25 $. The discrepancy mostly corresponds to 
sources with $z_{phot} \gtapprox 2$, but it must be noticed that the lensing inversion
technique is characterized by a trend against the identification of high-$z$
images, whereas $z_{phot}$ does not show this trend.
This behaviour is expected as a result of the relative low sensitivity to $z$ of the
inversion method for high-$z$ values. Thus, in general, the $z_{phot}$ determination is 
more accurate than the $z_{lens}$. Nevertheless, the combination of both distributions
is particularly useful when a degenerate solution appears using $z_{phot}$, 
as shown in Figure~2.

\begin{figure}
\psfig{figure=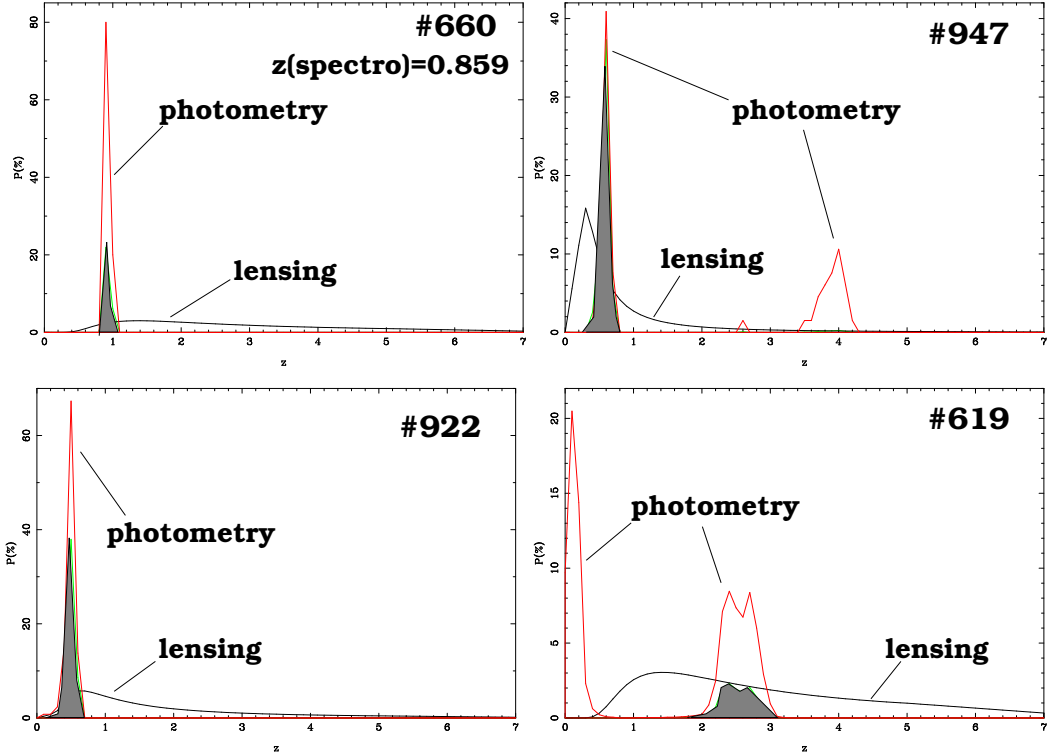,height=10.0cm}
\caption{Combined $z_{phot}$ and lensing inversion probability distributions for the redshift of
4 different sources in A2390, representing the typical cases found in this field. 
Shaded regions correspond to the final composite probability. For most objects,
such as $\#660$ and $\#922$, $z_{phot}$ gives a more accurate redshift value compared to
$z_{lens}$, except in the cases with a degenerate $z_{phot}$ solution ($\#947$ and $\#619$)
\label{fig2}}
\end{figure}

\section{Photometric redshifts and cluster lenses}

\subsection{Identification of multiple images}

Cluster lenses are useful only when their mass distribution is highly constrained
by multiple images, revealed by HST and multicolor photometry. The $z_{phot}$
technique is useful to identify objects with similar SEDs within the errors,
thus compatible with a multiple image configuration. $z_{phot}$s
can be used advantageously when data span a large wavelength range,
and particularly if near-IR photometry is available, because this allows to
obtain accurate $z_{phot}$ in the sensitive region of $0.8 \le z \le 2$.

\subsection{Scaling the mass in weak lensing analysis}

$z_{phot}$s are particularly useful when deriving the mass from weak shear
analysis: they are used to eliminate cluster and foreground
galaxies from the analysis, and to scale the lensing distance modulus in order
to compute the mass from the gravitational convergence.
The average convergence $\kappa \equiv \Sigma / \Sigma_{cr}$,
which corresponds to the ratio between the surface mass density and the critical
value for lensing, may be obtained as a function of the radial distance
$\theta$ using different methods (see Mellier 1999\cite{mel} for a review).
The mass within an aperture $\theta$ is given by
 
\begin{equation} 
M (<\theta)  = \kappa(<\theta) \ \Sigma_{cr} \cdot \pi \left(\theta D_{ol}\right)^2 \nonumber 
  =  \kappa\left(<\theta\right) \ \theta^2 \,\frac{c^2}{4G}
\left<\frac{D_{ls}}{D_{os}D_{ol}}\right>^{-1}
\end{equation} 

\noindent where $D_{ij}$ are the angular size distances between the cluster ($l$),
the observer ($o$) and the source ($s$), and $\kappa(<\theta)$ is the averaged convergence 
within the radius $\theta$. Using the $N(z_{phot})$ computed through a
suitable filter set, the mean value of 
$\left<\frac{D_{ls}}{D_{os}D_{ol}}\right>$ 
can be computed, thus leading to a fair estimate of the mass. This method has
been recently applied to the lensing clusters MS1008-1224\cite{Athreya} and 
A2219\cite{bez1} (see also Abdel Salam and Gray, this conference). 
Because of the small projected surface across the redshift space, the effective 
surface which is ``seen" through a cluster lens is relatively small, thus 
producing a strong variance from field to field. Obtaining the $N(z_{phot})$
distribution for each cluster could help to improve the mass determination.
Nevertheless, the distortion on the $N(z_{phot})$ distribution itself depends on
the mass distribution (sect. 2.1). An iterative process is needed to
reduce the error bar on the mass, which is $\sim 30 \%$ without this second order
correction\cite{Athreya}.

\subsection{Identification and Characterization of lensing structures}

$z_{phot}$s are useful to improve the detection of clusters in wide-field surveys, 
and to identify the visible counterpart of complex lenses.
It has been shown that including such a $z_{phot}$ technique in an automated identification 
algorithm allows to improve significantly the detection 
levels for clusters\cite{pel1}, whatever the algorithm used\cite{Kep} \cite{Ols} \cite{Kaw}. 
In general, the $S/N$ is improved by a factor of 3 to 6 up to $z \sim 1$,
depending on the redshift and magnitude limits. The detection efficiency 
in the $0.8 \le z \le 2.2$ region is improved only when using a $z_{phot}$ selection
based on optical {\it and} near-IR filters.
These comments also apply to multiple, complex and/or dark lenses,
where $z_{phot}$s allow to identify the main lensing structures. Examples of composite
lenses recently identified by $z_{phot}$ are MS1008-1224\cite{Athreya}, where a
secondary lens exists at $z \sim 0.9$, and the multiple-quasar fields 
Q2345+007\cite{pel3} and the Cloverleaf\cite{allo}.

Cluster members could be also selected using $z_{phot}$ criteria. The present version 
of {\it hyperz} is also able to display the probability of each object to be at a 
fixed redshift. This is useful when looking for clusters of galaxies at a given (or guessed)
redshift. In this way, the number density and luminosity density distributions of cluster galaxies
can be computed, and $M/L$ ratios could be estimated.

\section{Conclusions and Perspectives}

Concerning the lensing clusters, $z_{phot}$s appear as an essential tool for
mass calibration in weak shear studies (see the recent papers on MS1008 
\cite{Athreya} and A2219 \cite{bez1}). 
 $z_{phot}$s allow to improve the identification of the ``visible" counterpart
of lensing structures, to determine their redshifts and to measure $M/L$ ratios
of clusters, groups, etc.

   $z_{phot}$s allow to optimize the spectroscopic surveys of faint lensed sources
(visible versus near-IR domains), and to identify ambiguous spectral features (emission lines).
Before the recent VLT survey on AC114 (Le Borgne et al. this conference),
all spectroscopically confirmed lensed sources were ``bright" ($M_B \ltapprox -21$).
Because a redshift accuracy of $\delta z \sim 0.1 (1+z)$ is enough for most applications,
$z_{phot}$s allow to go further in magnitude when deriving the statistical properties of
faint sources. Besides, combining $z_{phot}$ and lensing inversion techniques
provides with an alternative way to determine the redshift distribution of sources.
A future development should be the study of the systematics and biases introduced 
by GTs when they are used to access the distant population of faint sources.
The typical uncertainty in the amplification factor is  $\Delta m \sim $
0.3 magnitudes, a value which is similar to model uncertainties when deriving 
intrinsic luminosities and SFRs of background sources with relatively well 
constrained SED ($\sim 30\%$ accuracy). Only  well known lenses are actually 
useful as GTs.

   Conversely, lensing clusters could be considered as a tool to calibrate $z_{phot}$s 
beyond the reach of standard spectroscopy, up to the faintest limits in magnitude,
thus allowing to extend the training set for $z_{phot}$s. 
The sample of lensing clusters available has to be
enlarged, in order to obtain a weakely biased image of the averaged
properties of sources along different lines-of-sight.
In order to sample a statistically significant field at $z \ge 2$ in the
strong amplification domain (close to the corresponding caustic lines),
we need to study about 10 different and well-known cluster lenses.
The Ultra-Deep Photometric 
Survey of cluster lenses and the subsequent spectroscopic follow up of sources
constitute a well defined program for 10m ground-based telescopes and the future NGST.

\section*{Acknowledgments}
This work was supported by the TMR {\it Lensnet} ERBFMRXCT97 - 0172
(http : // www.ast.cam. ac.uk /IoA/lensnet), the ECOS SUD Program,
the French {\it Centre National de la Recherche Scientifique},
and the  French {\it Programme National de Cosmologie} (PNC).


\section*{References}
\begin{thebibliography}{99}

\bibitem{Alt}  Altieri, B., et al. A \& A 343, L65 (1999).

\bibitem{Arn} Arnouts, S., Cristiani, S., Moscardini, L., Matarrese, S.,
Lucchin, F., Fontana, A., Giallongo, E., MNRAS 310, 540 (1999)

\bibitem{Athreya} Athreya R., Mellier Y., Van Waerbeke L., Fort B., Pell\'o R., Dantel-Fort M., A \& A
submitted, astro-ph/9909518 (2000).

\bibitem{bez1} B\'ezecourt J., Hoekstra H., Gray M. E. , AbdelSalam H. M. , Kuijken K. , Ellis R. S.,
A \& A submitted, astro-ph/0001513 (2000).

\bibitem{Blain} Blain, A. W., Kneib, J.-P., Ivison, R. J., Smail, I., ApJ 512, L87 (1999)

\bibitem{Bol} Bolzonella, M., Miralles, J.M., Pell\'o, R., A \& A {\it submitted}, astro-ph/0003380, (2000).

\bibitem{Ebb} Ebbels, T.M.D., et al., MNRAS 295, 75 (1998).

\bibitem{Fdez} Fern\'andez-Soto, A., Lanzetta, K.M., Yahil, A., ApJ 513, 34 (1999)

\bibitem{Franx} Franx, M.,  et al., ApJ 486, 75 (1997)

\bibitem{Fry} Frye, B., Broadhurst T., ApJ 499, 115 (1998)

\bibitem{Fur} Furusawa, H., Shimasaku, K., Doi, M., Okamura, S., ApJ in press, astro-ph/9912447, (2000)

\bibitem{Gia} Giallongo, E., D'Odorico,, S., Fontana, A., Cristiani, S.,
Egami, E., Hu, E., McMahon, R.G., AJ 115, 2169 (1998)

\bibitem{Gwy} Gwyn, S.D.J., Hartwick F.D.A., ApJ 468, L77 (1997)

\bibitem{Kaw} Kawasaki W., Shimasaku K., Doi M., Okamura S., A\&AS 130, 567 (1998)

\bibitem{allo} Kneib, J.-P., Alloin, D., Pell\'o, R., A \& A 339, 65 (1998).

\bibitem{Kep} Kepner J., Fan X., Bahcall N., Gunn J., Lupton R., ApJ 517, 78 (1999)

\bibitem{Lan} Lanzetta, K.M., Yahil, A., Fern\'andez-Soto, A., Nature 381, 759  (1996)

\bibitem{mel} Mellier, Y., ARA \& A 37, 127 (1999)

\bibitem{Mob} Mobasher, B., Rowan-Robinson, M., Georgakakis, A., Eaton, N., MNRAS 282, L7 (1996)

\bibitem{Ols} Olsen L.F., et al., A\&A 345, 363 (1999)

\bibitem{pel1} Pell\'o, R., Leborgne, J.F., Miralles, J.M., Bruzual, G., Proceedings of the 14th 
IAP Meeting "Wide Field Surveys in Cosmology" (1998).

\bibitem{pel2}  Pell\'o, R. et al., A \& A 346, 359, (1999).

\bibitem{pel3} Pell\'o, R., Miralles, J. M., Le Borgne, et al.,  A \& A 314, 73 (1996).

\bibitem{Smail} Smail, I., Ivison, R. J., Blain, A. W., Kneib, J.-P., AAS 192, 4813 (1998).

\bibitem{Saw} Sawicki, M.J., Lin, H., Yee, H.K.C., AJ 113, 1 (1997)

\bibitem{Tra} Trager, S. C., Faber, S. M., Dressler, A., Oemler, A. Jr., ApJ 485, 92 (1997).

\end{thebibliography}
\end{document}